\documentstyle[seceq,epsf,twoside]{ptptex}
\def \be {\begin{equation}}
\def \ba {\begin{eqnarray}}
\def \ee {\end{equation}}
\def \ea {\end{eqnarray}}

\def \qq {$q\bar q$}

\def \lsm {L$\sigma$M}
\def \sig {$\sigma$}

\def \be {\begin{equation}}
\def \ba {\begin{eqnarray}}
\def \ee {\end{equation}}
\def \ea {\end{eqnarray}}

\notypesetlogo  
\title{Summary talk of the conference on the sigma resonance \footnote{Invited talk at the Conference: ``Possible Existence of the Light $\sigma$ Resonance and
its Implications to Hadron Physics", Yukawa Institute for Theoretical Physics, Kyoto, Japan, 11-14th of June 2000.}}
\author{Nils A. T\"ornqvist1\footnote{email: Nils.Tornqvist@Helsinki.fi} }
\inst{Physics Department, 
POB 9, FIN--00014, University of Helsinki, Finland}
\date{August 10, 2000}
\abst{
This report summarizes the most important results presented at the recent conference held at the Yukawa Institute of Theoretical Physics in Kyoto, devoted to the recently confirmed light $\sigma$ resonance. Remarkably, all speakers at this meeting took the light \sig\ for granted and many mass and width estimations near 500 MeV 
were presented. We emphasize that if the light and broad \sig\ is accepted as a true resonance it explains many basic problems of low energy hadronic physics in a simple way, especially
if the linear sigma model (\lsm ) is used as an approximate effective low energy theory. }
\begin{document}
\maketitle

\setcounter{tocdepth}{4}
\section{Introduction and historical remarks}
This was a great meeting. I believe I never before learned so many new important things in such a short time of only 3 days. The large number of talks, 38, showed that light hadron spectroscopy,
especially the light $\sigma$ resonance together with confinement and breaking of chiral symmetry,  is again an important topic.

The \sig\ meson was introduced theoretically in connection with the linear sigma model\cite{lsm} (\lsm ) over 40 years ago. Looking back, already in the very first paper\cite{pdg1}  of what was to become the Review of Particle Physics of the Particle Data Group
there was an entry for a light ``$\sigma$", with a first reference to Samios et al.\cite{samios}. That \sig\  was certainly a statistical fluke, claiming  a narrow resonance peak of only 50 MeV and a mass of 395 MeV, i.e., certainly not the very ($\approx$500 MeV) broad \sig\ we discuss today\cite{pdg2000}. The same can be said to all later claims for a narrow sigma in the 500-700 MeV region, and many such candidates for a light and narrow \sig\ resonances remained in the PDG until 1972, when the
much discussed ``up-down" ambiguity of the phaseshift near 700 MeV was resolved in favour of the ``down" solution. This excluded
the possible existence of a narrow \sig . However, the possibility that there still could exist a 
very broad \sig\ resonance was not really excluded. But at that time most people did not want to believe in such a broad resonance, which could not easily  be distinguished from a background  without relying strongly on theoretical models.

So the light \sig\ disappeared from the tables for over 20 years, until the 1996 edition, when
Roos and I were able to convince the other members of the meson team (after we had convinced ourselves in our own analysis\cite{nat}) that we should introduce an
entry for a broad and light \sig\ resonance. Since then a large number of independent theoretical analyzes have confirmed that this was, in fact, the correct decision. The problem with the \sig\
is that in order to disentangle the resonance pole (especially from the $\pi\pi\to\pi\pi$ data, where a chiral zero supresses the low energy tail of the \sig ) one needs
elaborate models, which include in addition to unitarity, analyticity and coupled channels also chiral symmetry constraints and preferably an SU3 flavour symmetric framework. Experimentalists who analyze a broad structure like the \sig\ usually do not employ such elaborate framework.

Also  in the 1960'ies and early 1970'ies it was not well known that strong crossed channel effects should generally always be associated with resonance-like behaviour in the s-channel. This obscured the interpretation of the \sig\ as a broad resonance and it was believed that the sigma
``bumps" were just background from strong t-channel exchanges. Later, when the duality of Regge exchanges and resonances (the Veneziano model) became widely accepted, it was well understood that, on the contrary, strong crossed channel effects should 
always come together with a resonance. Thus although $\rho$ etc. exchange can, within a specific model, generate a large $\pi\pi$ phase shift, it does not disprove a resonance interpretation, but instead strengthens it (see also Igi\cite{conf}). 
But, before that was accepted,  the light sigma had been replaced by
a heavier and narrower candidate, which from 1972 to 1996 entered under different  names like ``$\epsilon$(1200)" or $f_0$(1400). 
\section{The mass and width parameters of the \sig\ reported at this meeting.}
In Fig.1 I have plotted with filled circles the results of 22 different analyses on the \sig\
pole position, which are included in the  2000 edition of the Review of Particle Physics\cite{pdg2000} under the entry $f_0(400-1200)$ or \sig .  Most of these
find a \sig\ pole position near 500-i250 MeV. 
 At this meeting we have heard
of many preliminary analyzes, which also find \sig\ resonance parameters in the same region. I have listed these in Table 1 and plotted them with triangles in Fig.1. 
It was not possible for me to distinguish between Breit-Wigner
parameters and pole positions, which of course can differ by several 100 MeV for the same data
(see also discussion in Sec. 4). Also it must be noted that many of the triangles in Fig.1 rely on the same raw data and come from preliminary analyzes not yet published.
\begin{table} 
\caption{Values of $\sigma$ resonance parameters 
reported in the various talks at this meeting.}
 \begin{center}
 \begin{tabular}{|c|c|c|c|}
 \hline 
 Speaker\cite{conf}  & Comment & Mass (MeV)    & $\Gamma_{\pi\pi}$ (MeV)
 \\
 \hline
 Takamatsu& Central prod.   & $590\pm 10$ &$710\pm 30$      \\  
 Takamatsu& E135   & $588\pm 12$ &$618\pm 25$      \\  
 Harada&  $\pi\pi\to\pi\pi$    & $559$ &$378$      \\  
 Kaminski & $\pi\pi\to\pi\pi$   & $520$ &$520$      \\  
 M. Ishida & $\pi\pi\to\pi\pi$   & $535\sim 675$ &$385\pm 70$      \\  
 Shabalin& $\pi\pi\to\pi\pi$& $700$ &$725$      \\  
 Tsuru& Central prod. 4$\pi$           & $450\pm 50$ &$600\pm 50$     
\\  
 Tsuru& DM2 $J/\psi\to \omega\pi\pi$   & $414\pm 20$ &$495\pm58$      \\
  Komada&  $\Upsilon'\to \Upsilon\pi\pi$   & $521$ &$316$      \\  
 Kobayashi& GAMS $3\pi,\eta\pi$   & $560 ?$ &$403 ?$      \\  
\hline  
\end{tabular} 
\end{center} 
\end{table}
I also included in Fig. 1 with a star the \sig\ parameters obtained from
the recent E791 experiment at Fermi lab\cite{E791}, where 46 per cent
of the $D^+\to 3\pi$ is $\sigma\pi$.

\begin{figure}
\epsfxsize=14 cm
\epsfysize=6 cm
\centerline{ \epsffile{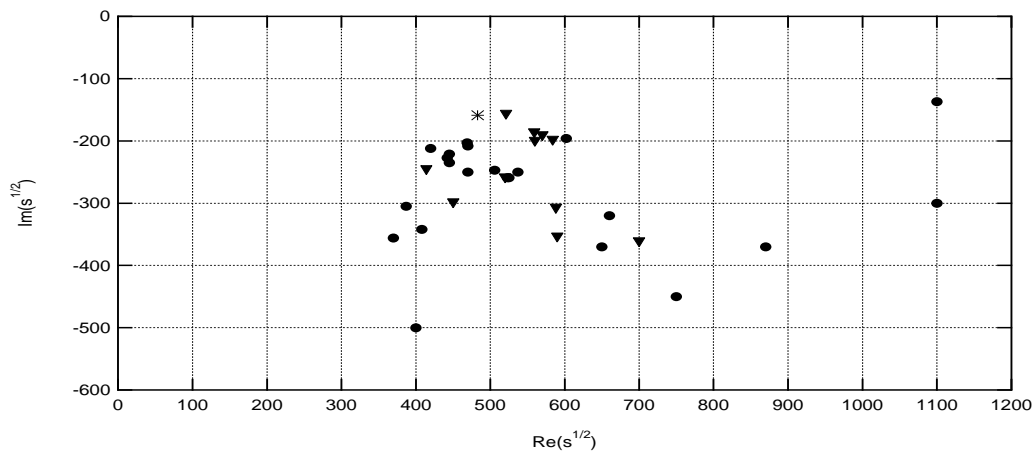}}
 \caption{ The pole positions of the \sig\ resonance, as listed by the PDG\cite{pdg2000} under $f_0(400-1200)$ or \sig\ (filled circles), plotted in the complex energy plane (in units of MeV).
The triangles represent the mass and width parameters (plotted as $m-i\Gamma /2$), which were reported at this meeting.
I could not here distinguish between pole and Breit-Wigner parameters. The star
is the $m-i\Gamma /2$ point obtained from the recent E791 experiment\cite{E791} on $D\to\sigma \pi\to 3\pi$ ($m_\sigma=478$ MeV $\Gamma_\sigma = 324$ MeV).}
\end{figure}

From that same experiment E791 we heard a rumour that they  seem to see a light 
$\kappa$ in $D^+\to K^-\pi^+\pi^+$. That would make some of us happy (Harada, Ishida, Oller,  Scadron,  Takamatsu, Tsuru ... )\cite{conf}, but I really believe that at this point we must remain a little skeptical, since if the LASS experimental $K\pi$ (nearly elastic) S-wave phase is right, it shows  only a very smooth behaviour up to the $K\eta'$ threshold and passes $90^\circ$ only at 1350 MeV. This is  different from the situation with $\pi\pi$ phase shift, where the $90^\circ$ value is reached at 700-800 MeV, not too far from the \sig (500-600). A very large
destructively interfering background could of course shift the  resonance parameters, from naive expectations, but that is very sensitive to how unitarity+analyticity is imposed.
Therefore, it remains as an interesting task for the proponents of a light and broad $\kappa$ to show how the very general arguments of Cherry and Pennington, ``There is no light kappa"\cite{cherry}, can be circumvented. I remain with an open mind in this question, although I tend to agree with 
Cherry and Pennington, because in my analysis\cite{nat,conf} I can easily understand why the strange scalar is so much heavier than the others. Possibly I could understand such a light
$\kappa$ if this is related to the resonance doubling phenomenon, which we discussed with Roos\cite{nat}.

I found it remarkable that at this meeting there was not a single talk, which clearly doubted 
the existence of the \sig ! Every speaker took the \sig\ for granted. Was this because noone dared to, in front of this audience, express an opposing view? Or are we witnessing
a ``bandwagon effect", where everyone wants to join the club. I would rather guess that we are
observing a minor ``shift of paradigm", where more and more people are realizing that with the light \sig\ many things, from chiral symmetry breaking, constituent quark mass generation, $\Delta I=1/2$ rule  to perhaps even confinement, become much  easier to understand.   
\section{Personal recollection from the 38 talks}
At this meeting we learned and were reminded of many interesting things.
The linear sigma model (L$\sigma$M) was mentioned by many speakers (M. Ishida, Kunihiro, Scadron, Shabalin, T\"ornqvist)\cite{conf}, and we were honored to have Joe Schechter among us, one the originators of the $U3\times U3$ \lsm . 

Mike Scadron\cite{conf} showed how starting from the $SU2\times SU2$ \lsm\ with a quark-meson coupling $g$, (for which he finds $g= 2\pi/(N_c)^{\frac 1 2 }=3.63$), one obtains $f_\pi\approx 90 $ MeV, $m_q\approx 325$ MeV, and the Nambu relation $m_\sigma =2m_q\approx 650$ MeV. His inputs
are (apart from that in the chiral limit $m_\pi$ vanishes and that the Goldberger-Treimam  relation for quarks $m_q=f_\pi g$ holds) self-consistency conditions involving  one-loop gap equations. These require
a logarithmic cutoff at about 750 MeV. Then calculating further quark loop diagrams he predicts tri- and four-linear couplings between the $\pi$'s and $\sigma$'s. One important observation in his scheme is that quadratic divergences are softened to logarithmic
ones, by noting that with dimensional regularization (or with any  conventional reasonable regularization) one finds that the massless tadpole (which naively diverges as $\Lambda^2$), in fact, must vanish. This  leaves only one logarithmic cut-off parameter in agreement with Veltman's condition. This seems to be a simple way to get similar predictions
as one gets by bosonizing the NJL model. 

 About  30 years ago Schechter and Ueda\cite{lsm}  wrote down the $U3\times U3$ \lsm\ for the meson sector involving
a scalar and a pseudoscalar nonet. This (renormalizable) theory has only 6 parameters, out of which 5 can be fixed by the pseudoscalar masses and decay constants ($m_\pi,\ m_K, \ m_\eta,\ f_\pi, \ f_K$). The sixth parameter for the OZI rule violating 4-point coupling must be small. One can then predict with no free parameters the tree level scalar masses\cite{testa} (See Table 2), which turn out to be not far from the lightest experimental masses, although the two quantities are \underline{not} exactly the same as we stress in Sec. 4. 
\begin{table}
\caption{Tree level scalar masses predicted from the $U3\times U3$ \lsm\ at the tree level, when the parameters are fixed from pseudoscalar data, and experimental real part of pole positions}
\begin{center}
\begin{tabular}{|c|c|c|c|c|}
\hline 
&$m_\sigma $ &$ m_{f_0}$ & $m_\kappa $ &$ m_{a_0}$  \\
\hline
Tree level \lsm   & 620&1190&1120&1128\\
\hline
Experiment\cite{pdg2000}& $\approx$600&980&1430&988\\
\hline 
\end{tabular}
\end{center}
\end{table}
The important thing is that the scalar masses are predicted near the lightest experimentally seen scalar masses, and not in the 1500 MeV region where many authors
want to put the lightest $q\bar q$ scalars. Also the \sig\ is predicted to have a very large width ($\approx 500$ MeV), which agrees with our present understanding of the data.

Kunihiro\cite{conf} recalled the old problem of Gasser and Leutwyler\cite{gass} that when comparing the \lsm\ with the $L_i$
parameters of chiral perturbation theory ($\chi PT$), they noted by   that $L_6$ is predicted to be 6 compared to the
experimental $16.5\pm1.5$. But this does not necessarily deny the linear
realization of chiral symmetry as given in NJL-like models.
Shabalin  emphasized later, that one should not compare with too naive \lsm 's and that if one includes smooth phenomenological form factors  in the vertex functions (instead of sharp cut-offs) good agreement with data is obtained. Such form factors are certainly very physical and can be related to the fact that hadrons are not point-like particles as in naive field-theoretic models, but certainly have a finite size of 0.6-0.7 fm. Anyone who has worked on hadronic widths in e.g. the $^3P_0$ model, or unitarized  quark models, like myself,  are familiar with such
phenomenological form factors. Thus, effective low energy field theories,
 should be cut-off smoothly through form factors,
which certainly change their predictions as one approaches the high energy end of their region of applicability.

Kunihiro also mentioned that the $\pi N$ sigma term gets 
enhancements from the $\sigma(\approx 600)$ and that one predicts $\Sigma_{\pi N}=\hat m<u\bar u +d\bar d >= 49$ MeV,
which is near the 45 MeV ($\pm 20$\% ) determined from experiment\cite{sainio}, without too much
$s\bar s$ components in the nucleon wave function.

The $\Delta I=1/2$ rule in $K_S\to\pi\pi$ was discussed by several speakers
(Kiyo, Kunihiro, Sanda, Scadron, Shabalin)\cite{conf} and that  it gets a very simple
explanation once one accepts a $\sigma$ near 500 MeV. 
The final state interaction with the $\sigma$ resonance simply enhances
the I=0 final state. 
Kiyo also discussed the $\epsilon'/\epsilon$ parameterin K decays and
Terasaki discussed the \sig\ contribution 
 to the $K_L-K_S$ mass splitting, and both concluded that \sig\ contributes, but is not enough to explain the data. In a recent paper\cite{polosa} also other evidence for the
\sig\ resonance in weak interactions were discussed. 

We saw new $\pi^0\pi^0$ phase shift data from KEK Benkei E135 experiment shown by Takamatsu\cite{conf}.
These showed, what most people had expected,  that the old Cason data give  too low phase shifts. The new data agree with the main ``down-flat" solution of the Krakow group showed by Kaminski. But these data have still to be corrected for $a_1$ exchange in the production.

Kaminski also discussed the Roy equations as did T. Sawada and Shabalin. Here we all wait for the new data expected from the DIRAC and KLOE experiments.  T. Sawada showed in his analysis of the Roy equations an exciting
discrepancy, which may require a new long range Van der Waals interaction in $\pi\pi$ interactions! Could this be due to anomalously large 2$\gamma$ exchange or 
massless two-gluon exchange?

S. Sawada and Tamagaki\cite{conf} reviewed the old problem of how \sig\ and $2\pi$ exchange is needed in the understanding of nuclear binding and NN interactions. Tamagaki pointed out that maybe
experiments with $\Lambda\Lambda$ hypernuclei might be able to resolve whether \sig\ or $2\pi$ exchange 
(with intermediate $\Delta(1232)$\cite{green} or hyperons) is the dominating effect.

S. and M. Ishida\cite{conf} reminded us that the Adler zeroes, which suppress $\pi\pi\to\pi\pi$ low energy interactions, because of the chiral cancelation of \sig\ production and background,
are absent in production processes. (Therefore  the phenomenological $\alpha(s)$ functions
used  by  Au, Morgan and Pennington\cite{au} for production processes, are not smooth, but must cancel the Adler zeroes.) Consequently the \sig\ stands up more clearly
in production processes like central production $pp \to p_f (\pi\pi) p_b$, $\Upsilon\to\omega\pi\pi$, $\Upsilon'\to\Upsilon\pi\pi$ etc. We  saw some beautiful fits with \sig\ in many of these processes, which often previously were fitted using a broad background instead of the \sig . See the talks\cite{conf} by  Kamada, Kobayashi, M. Ishida and Tsuru. Also the exciting recent measurement of $D\to\sigma\pi\to 3\pi$ by the E791 experiment belongs to this cathegory.
A prediction of the $D^+\to\sigma \pi^+$ rate has  recently been calculated by Gatto et al.\cite{gatto} in agreement with this new data.

S. Ishida\cite{conf} argued for unconventional relativistic S-wave \qq\ bound states with $0^{++}$ quantum numbers (i.e. not as in the usual quark model),
which he and his collaborators call ``chiralons" in their ``covariant oscillator quark model". These would appear only at low masses and would include the lightest scalars (\sig , $\kappa ,\ a_0,\ f_0$), one of the iotas and a possible new light $a_1$ candidate. This is of course still very speculative, but if this has some  truth, it would certainly open up a new approach.

Oller showed many beautiful fits with the \sig\ obtained by starting using chiral perturbation theory  and iterating loops to generate bound states for \sig , $\kappa,\ a_0(980)$ and $f_0(980)$. In this way one generates a light \sig , although, as in the nonlinear \sig\ model,  the \sig\ is absent at the tree level. The convergence radius (which naively without loops is at the first threshold\cite{truong}) certainly increases considerably this way, and their analysis  agrees with all others in that once one imposes chiral symmetry constraints with unitarity, analyticity, flavour symmetry etc. one necessarily finds the light and broad \sig\ pole. Thus by including a 
Dyson sum of loop diagrams in the inverse propagators, as Oller and collaborators do, $\chi PT$  predicts the \sig . 

Certainly, also  the \lsm\ must be looked upon as an effective theory, and also has a limited  region of validity, extending at most to a little above 1 GeV for the S-wave mass spectrum.
It is a very instructive model, which is easy to understand. It includes the scalars and pseudoscalars (and unconfined quarks) from the start (with mesons as CDD poles like \qq\ states), and provides in principle a  formalism of how to include and calculate unitarity, analyticity, flavour, chiral and even crossing symmetry constraints in a well defined framework. It is even a renormalizable theory, although that property is not very essential in practice, since we know it can at most be a reasonable approximation to the true theory at low energy. 
I used to say ``The difficulty of physics is to see its simplicity" and personally I think that if a simple model explains most of what you see, then study that simple model first before you look for more elaborate theories.  Therefore
I hope that more people would spend some more time in studying the \lsm\ and its generalizations in more detail. E.g. to my knowledge very few authors have gone beyond the tree level and really calculated, say the pole positions of the scalar nonet in the $U3\times U3$ meson model.      

K. Igi\cite{conf,igi} showed using a framework including crossing symmetry and $\rho$ exchange together with untarity and analyticity it is not possible to understand the $\pi\pi$
phase shift without the \sig\ pole. Only half of the rise in the S-wave phase shift could be generated from $\rho$ etc. exchanges. This supports M. Roos' and my analysis\cite{nat} and shows that the critical comment by Isgur and Speth\cite{isgur} can not kill the \sig .

Different unitarizations were discussed by many speakers\cite{conf} (M. and S. Ishida, Kaminski, Obu, Oller, Schechter,  Shabalin, T\"ornqvist).
These include first unitarizing the background and resonance separately and then multiplying the two S-matrices (or adding the two phases in the case of a single channel), N/D or  K-matrix unitarization. Now, unitarizing is in principle simple and can be done in many different ways. Its more difficult to have
``correct" physical analyticity especially with coupled channels. For the latter dispersion relations usually help a lot to e.g. 
avoid spurious singularities, which otherwise easily creep into the amplitudes.
In spite of the old nature of this subject, I believe we still have quite some work to do, in order to sum approximately   the infinite number of loop diagrams, which Nature does in a unique way.

Finally today we heard Hatsuda, Shimizu, Hirenzaki, Ozawa and Iwasaki\cite{conf} discuss how to measure
the mass and width parameters of \sig , $\omega,\ \phi, \ ...$ in a nuclear medium. It certainly is an exciting possibility if we in the future can study  if and how the mass and width of the \sig\ decreases as it is produced inside a nuclear environment. Those are difficult experiments and we must wish them the best of luck. The last five talks by Lim,
Takizawa, Nishiyama, Matsuki and Watabe\cite{conf}, just before my summary talk, were of 
rather theoretical nature, and I did not have time nor energy to comment on them. 
\section{Concluding remarks}
Finally let me remind you again of the important differences in different mass and width definitions,
which appear in the literature. Already in the simple case of one decay channel and one
resonance one can define these in different ways. Recall that the inverse propagator is 
\be P^{-1}(s) = 
m_0^2+{\rm Re}\Pi (s) +i {\rm Im}\Pi (s) -s 
\ee
Then the following three mass (and width)  definitions are different, and  for the same data for a broad resonance one can get values 100's of MeV apart:
\begin{enumerate}
\item[({\it i})] {Lagrangian mass parameters, where one essentially neglects the 
 $\Pi (s)$  or assumes it is a real renormalization constant, which can be included in $m_0^2$. The latter  then defines the tree level mass. }
\item[{(\it ii})]{Breit-Wigner or 90$^\circ$ mass, where one looks for the zero in the real expression $m_0^2+{\rm Re}\Pi (s) -s$, and often  also uses an approximation that ${\rm Re}\Pi (s)$ is a constant. The BW mass is then defined through that zero, and the width is taken as
$\Gamma ={\rm Im}\Pi (s=m^2)/m$.}
\item[({\it iii})]{Second sheet pole position parameterized as $m-i\Gamma /2$. Here one looks for the nearest zero in the complex energy plane (with the unitarity cuts) of the whole expression above. This is of course the best, since only at the pole can one
factorize the production and decay. Only this mass is independent of the reaction. The difficulty is that one must have a very good theory.
 }\end{enumerate}
For example in Table II the theoretical values are as in ({\it i}), while the experimental masses
are as in ({\it ii}) or ({\it iii}) above.
For the \sig\ it is especially important to have a good theory, since it is so broad. This means that apart from a ``correct" unitarity, analyticity and coupled channel framework one needs constraints from a chiral, flavour and even crossing symmetries. 
I believe the \lsm\ is a good starting point, but we need better models.

The \sig\ meson is perhaps the most important meson in the book. 
In Swedish and Finnish we have a saying ``A loved child has many names" (Ett k\"art barn har m\aa nga namn). The \sig\  certainly has many names:
\begin{itemize}
\item{$f_0(400-1200)$ or \sig\ (PDG)\cite{pdg2000}.}
\item{Quantum fluctuation of the chiral order parameter (Hatsuda, Kunihiro)\cite{conf}.}
\item{Chiralon (S. and M. Ishida)\cite{conf}.}
\item{$\pi\pi$ correlation or bound state (Oller\cite{conf}).}
\item{Gluonium-\qq\ mixed state (Narison\cite{conf}).}
\item{4-quark state (Schechter\cite{conf}, R. Jaffe\cite{jaffe}).}
\item{The eye witness of confinement (Gribov\cite{gribov}, originally for $f_0(980)$ and $a_0(980)$, since the \sig\ was then absent in the PDG).}
\item{The sister particle of Yukawa's pion\cite{conf}.}
\item{The Higgs of strong interactions\cite{conf}, giving the light constituent quarks most of their mass.}
\end{itemize}

The name $f_0(400-1200)$ of the PDG looks like a name for a ``junk entry", usually used for
a collection of miscellaneous peaks, bumps or resonances seen by  various groups, which are  questionable and noone  understands. Looking at fig.1 I think it should be renamed $f_0(500)$ or $f_0(600)$.
\acknowledgements
We  thank all the organizers for organizing this exciting meeting on the \sig\ resonance, which
I hope will be the beginning of a new series of meetings. I also thank the organizers for inviting me to give this summary talk. I was afraid this would give a lot of hard work, but in fact, I enjoyed it as it forced me to really try to understand the essential things in every talk. Thus many thanks go to:
Teiji Kunihiro, Tetsuo Hatsuda, Tsuneaki Tsuru, Kunio Takamatsu,  Chong-Sa Lim, Hajime Shimizu,
Makoto Oka, Shin Ishida, Muneyuki Ishida. I also thank A. D. Polosa for commenting on the manuscript.

\end{document}